\documentclass[10pt,twocolumn]{IEEEtran}

\usepackage{epsf,psfig}
\usepackage{amsmath, amsthm, amssymb}
\newtheorem{theorem}{Theorem}[section]

\begin{document}

\title{On the Unicast Capacity of Stationary Multi-channel Multi-radio Wireless Networks: Separability and Multi-channel Routing}
\author{Liangping~Ma 
\thanks{Liangping Ma's work was supported in part by NSF grant CNS-0721230. E-mail: liangping.ma@ieee.org.}}

\maketitle

\begin{abstract}
The first result is on the separability of the unicast capacity of stationary
multi-channel multi-radio wireless networks, i.e., whether the capacity of such a network is equal to the sum of
the capacities of the corresponding single-channel single-radio
wireless networks. For both the Arbitrary Network model and the Random Network model, given a channel assignment, the separability property does not always hold. However, if the number of radio
interfaces at each node is equal to the number of channels, the
separability property holds. The second result is on the impact of multi-channel routing (i.e.,
routing a bit through multiple channels as opposed to through a
single channel) on the network capacity. For both network models,
the network capacities conditioned on a channel assignment under the two routing schemes are not always equal, but if again the number of radio interfaces at each node is equal to
the number of channels, the two routing schemes yield equal network capacities.
\end{abstract}


\section{Introduction}
It is well known that the joint capacity of independent parallel
Gaussian channels with a total power constraint is equal to the
sum of the capacities of the individual channels~\cite{Cover06}.
Therefore, to find the joint capacity of the parallel channels, we
can find the capacities of the individual channels separately, and
then sum them up. We call such property \emph{separability} in
channels. The separability property may look trivial but it is not.
In fact, when the noise is colored, i.e., dependent from
channel to channel, the separability property in general does not hold~\cite{Cover06}.

In this paper, we first study an analogue of the above separability property for stationary multi-channel multi-radio wireless networks. Specifically, given a multi-channel multi-radio network, we define the corresponding
single-channel single-radio wireless networks, and examine whether the capacity of the multi-channel multi-radio
network is equal to the sum of the capacities of those corresponding single-channel
single-radio networks.

We then investigate the impact of multi-channel routing on the
network capacity. In a multi-channel multi-radio network, two
routing schemes can be adopted: routing a given bit either (1) on
multiple channels, or (2) on only one channel while different bits
may be routed through different channels. We refer to the first
scheme as \emph{multi-channel routing}, and the second as
\emph{single-channel routing}. As an example, consider the routing
of bit $b$ from source (node $A_1$), through relays (nodes $A_2$ and $A_3$), to destination (node $A_4$). When multi-channel routing is adopted, a route may look like:
\begin{equation}
\mbox{ $A_1$ $ \stackrel{\mbox{channel 3}} {\longrightarrow} $
$A_2$ $ \stackrel{\mbox{channel 1}} {\longrightarrow} $
$A_3$ $ \stackrel{\mbox{channel 2}} {\longrightarrow} $ $A_4.$} \nonumber
\end{equation}
In contrast,
when single-channel routing is adopted, a route may look like:
\begin{equation}
\mbox{ $A_1$ $ \stackrel{\mbox{channel 2}} {\longrightarrow} $
$A_2$ $ \stackrel{\mbox{channel 2}} {\longrightarrow} $
$A_3$ $ \stackrel{\mbox{channel 2}} {\longrightarrow} $ $A_4$,} \nonumber
\end{equation}
while a different bit $b'$ may be routed through a different channel, say, channel 3. Note that by definition multi-channel routing includes single-channel routing as a special case.

Two network models are considered: Arbitrary Network and Random Network~\cite{Kumar00}. The communication links are point-to-point with fixed data rates, and advanced techniques such as successive interference cancelation or MIMO~\cite{Tse_book05} are not considered. We assume that each node has $m$ radio interfaces, and there are $c$ orthogonal channels. Due to the assumed communication model, there is no benefit for a node to simultaneously transmit (or simultaneously receive) on multiple radio interfaces on the same channel. As a result, the case $m > c$ reduces to the case $m=c$, and therefore we consider only the case $m \leq c$.

The main results are as follows:
\begin{enumerate}
\item For both Arbitrary Networks and Random Networks, given a channel assignment, the separability property of network capacity does not always hold. But if $m=c$, the separability property holds.
\item For both network models, multi-channel routing in general yields equal or a higher network capacity than single-channel routing does. But if $m=c$, the two routing schemes result in equal network capacities.
\end{enumerate}

A striking difference of this paper from most existing work ~\cite{Kumar00}\cite{Kya05}\cite{Ma07bound}\cite{Fran07}\cite{Had07} is that this paper deals with the network capacity and not the bounds of it. Those bounds, although very useful for studying the asymptotic or rough behavior of large
wireless networks, are insufficient for studying the precise relations (i.e., being equal or unequal) to be evaluated in this paper.

The remainder of this paper is organized as follows. Section
\ref{sec_not} introduces some key notations, Section \ref{sec_arb} and
Section \ref{sec_rand} present the results for Arbitrary Networks
and Random Networks, respectively, and Section \ref{sec_implications} points out some implications of the results.

\section{Common Notations}\label{sec_not}
A \emph{multi-channel multi-radio network} $\mathcal{N}$ is a
4-tuple $(U,H,\sigma, \eta)$, where
\begin{itemize}
    \item $U:=\{1,...,n\}$ is a set of $n$ nodes, where
    each node $i \in U$ has $m$ radio interfaces,
    \item $H:=\{1,...,c\}$ is a set of $c$ channels, where channel $i \in H$ supports a fixed data rate of $w_i$ bits/sec,
    \item $\sigma$ is the region in which the nodes are located, and
    \item $\eta$ is the interference model.
\end{itemize}

A channel assignment distributes the $mn$ radio interfaces onto
the $c$ channels. Let $I_i$ be the set of radio interfaces of
$\mathcal{N}$ assigned to channel $i$, $i=1,...c$. For the network
$\mathcal{N}$ defined above, we define $c$ corresponding
\emph{single-channel single-radio networks}
$\mathcal{N}_i':=(I_i, i,\sigma, \eta)$, $i=1,...,c$. That
is, $\mathcal{N}_i'$ consists of the radio interfaces assigned to
channel $i$, and they can communicate only on channel $i$.

A network can be configured in different ways, resulting in
different data delivery capabilities. Define a \emph{network
configuration} $G$ as a 5-tuple $(X, I, F, M, P)$ for network $\mathcal{N}$,
where
\begin{itemize}
    \item $ X = (X_1, ..., X_n)$ is the locations of the
    nodes, where $X_i$ is the location of node $i$.
    \item $I=(I_1, ..., I_c)$ denotes the channel
assignment.
    \item $F = (F_1,
..., F_n)$ denotes the traffic flow configuration, where $F_i$ specifies a
traffic flow originating from node $i$.
 \item $M$ denotes the routing scheme, specifying the route for any source-destination pair.
 \item $P = (P_{ij})$, $i=1,...n,j=1,2,...$ is the transmission power
configuration, where $P_{ij}$ specifies the transmission power of the $j$th
transmission from node $i$.
\end{itemize}
Note that when $m=c$, the optimal channel assignment $I$
is simple: a 1-1 mapping between a node's interfaces and the
channels.

\section{Results for Arbitrary Networks}\label{sec_arb}

In the Arbitrary Network model~\cite{Kumar00}, node locations are
arbitrary, each node arbitrarily chooses a destination, and the
power level of each transmission is set arbitrarily. The network
capacity is measured by transport capacity~\cite{Kumar00}.
Following the convention in the communication theory that the term
``capacity'' refers to the supremum of a set of achievable
``rates'', we define transport capacity as the supremum of
achievable transport rates, which is defined below.

The \emph{unicast transport rate} of network $\mathcal{N}$ under
configuration $G$ during time interval $T$ is defined as
\begin{equation}
R(G,T) := \frac{1}{T}\sum_{b: b\in \langle T \rangle} l_b(G),
\label{eq_R1}
\end{equation}
with unit bit-meters per second, where $l_b(G)$ is the distance
(magnitude of the displacement) that bit $b$ travels from source to destination under
configuration $G$, and $\langle T \rangle$ is the set of bits
delivered within time interval $T$.

Assume that the diameter of the region
$\sigma$ and the data rates $w_i$ are bounded. Then, transport
rate $R(G, T)$ is also bounded and has a unique supremum, which we
define as the \emph{unicast transport capacity}
\begin{equation}
C(T) := \sup_{G} ~~R(G,T). \label{eq_cap1}
\end{equation}

Likewise, we define the unicast transport capacity for a
single-channel single-radio network $\mathcal{N}_i'$ in time
interval $T$
\begin{equation}
C_i'(T) := \sup_{G_i'} R(G_i', T),
\end{equation}
where $G_i'$ configures the interfaces assigned to channel $i$ under $G$. We can also define various conditional transport capacities. For example, the network capacity conditioned on a given channel assignment $I$ is define as
\begin{equation}
C(T|I) := \sup_{G | I } ~~R(G,T).
\end{equation}
where $G|I$ means that the channel assignment component of $G$ is fixed at $I$.

A few more definitions:
\begin{itemize}
    \item A
\emph{tick} $\tau_i$ is the time required to transmit 1 bit by one
hop on channel $i$, i.e., $\tau_i = 1/w_i$.
    \item A
\emph{Simultaneous Transmission Set (STS)} of channel $i$ is a set
of successful one-hop transmissions on channel $i$ in a tick.
    \item
Network $\mathcal{N}^a$ is said to \emph{simulate} network
$\mathcal{N}^b$, if there is a way for $\mathcal{N}^a$ to
replicate the delivery of all the bits delivered by
$\mathcal{N}^b$. The technique of simulation has been used in other places such as
proving the equivalence of the Deterministic Finite Automaton
(DFA) and the Nondeterministic Finite Automaton
(NFA)~\cite{Sipser06}. We should distinguish the simulation here
from what is performed by the network simulators such as the
Network Simulator-2 (NS-2)~\cite{ns}. With NS-2, a single computer
simulates the events that occurred in a computer network, but it
does not replicate real communication.
\end{itemize}

 Now we present the first result Theorem \ref{theorem_arb_sep1} below. To understand it, note that finding the transport capacity conditioned on a channel assignment involves optimization over various configuration parameters including the locations of nodes or radio interfaces. In a multi-channel multi-radio network, all interfaces of the same node must take the same location, and we call this constraint the \emph{interface location constraint}. In contrast, in the corresponding single-channel single-radio networks, which are optimized independently, there is no such constraint, therefore potentially resulting in a different conditional transport capacity. 
\begin{theorem}
For a multi-channel multi-radio Arbitrary Network, given the channel assignment, the transport capacity is not always separable in channels.\label{theorem_arb_sep1} \end{theorem}
\noindent \textbf{Proof:} We prove it by showing that there exists
a multi-channel multi-radio network $\mathcal{N}$ whose transport capacity conditioned on a channel assignment $I$ satisfies $C(T|I) <
\sum_{i=1}^c C_i'(T|I)$, where $C_i'(T|I)$ are the conditional transport capacities of the corresponding single-channel single-radio networks $\mathcal{N}_i'$. The network region $\sigma$ is the closure
of a $1$ meter $\times$ $1$ meter square, $n=4$, $m=2$, $c=3$,
$w_i = 1$ bits/sec $\forall i$, and the channel assignment is $I= (\{1,2,3,4\}, \{1,2\},
\{3,4\})$, which assigns 4 interfaces to channel 1, and 2 interfaces to each of the remaining two channels. The interference model
$\eta$ is the Protocol Model~\cite{Kumar00}, which states that a
transmission from node $i$ to node $j$ over some channel is
successful if $|X_k - X_i| \geq (1+\Delta)|X_i-X_j|$ for any other
node $k$ that simultaneously transmits on the same channel.

Let the optimal configuration conditioned on $I$ of network $\mathcal{N}$ be $G^*$. Since the nodes are distributed over 3 channels, there must be 3 or 4 simultaneous transmissions in $\mathcal{N}$ in any tick under $G^*$: 1 or 2 simultaneous transmissions on channel 1, and 1 transmission on channel 2 and channel 3
each.

For network  $\mathcal{N}$, the optimal routing conditioned on channel assignment $I$, denoted by $M^*$ is such that
each flow consists of only one hop. This way, the distance of each transmission is fully accounted in the transport rate while the maximum number of simultaneous transmissions can be achieved. As a result, $C(T|I)=R(G^*,T)=\sum_{i=1}^3 R_i(G^*,T)$, where $R_i(G^*,T):=\frac{1}{T}\sum_{b\in \langle T, i \rangle}
l_b(G^*)$ with $\langle T, i \rangle$ defined as the set of bits
delivered on channel $i$ in $T$. By the definition of $C_i'(T|I)$, we have
$R_i(G^*,T) \leq C_i'(T|I)$, and hence
\begin{equation}
C(T|I) \leq \sum_{i=1}^3
C_i'(T|I).\label{eq_proof1}
\end{equation}

Now consider the optimal configurations $G_i'^*$ of the corresponding single-channel
single-radio networks $\mathcal{N}_i'$, $i=1,2,3$. It is clear
that $G_2'^*$ is to place nodes 1 and 2 at the opposite ends of a
diagonal of square $\sigma$ and let one node transmit a time,
and $G_3'^*$ is similar. To achieve the equality in $\leq $ in
(\ref{eq_proof1}), we must have $R_i(G^*,T) = C_i'(T)$ for
all $i=1,2,3$, which, however, is impossible for some $\Delta$ as
shown next.

Suppose that $G^*$ satisfies $R_i(G^*,T) = C_i'(T|I)$ for $i=2,3$. Under $G^*$, the
interfaces in network $\mathcal{N}$ assigned to channels 2 and 3 must be
at the corners of square $\sigma$. Because those interfaces come
from all four nodes, the remaining interfaces, which are on
channel 1, will also be at the corners. Choose $\Delta > 0$ in the Protocol Model such that $G_1'^*$ allows two simultaneous transmissions on channel 1. It can be checked that placing all nodes at the corners of square $\sigma$ violates the constraints imposed by the Protocol Model and therefore cannot be optimal for $\mathcal{N}_1'$. Thus $R_1(G^*, T) < C_1'(T|I)$. By (\ref{eq_proof1}), $ C(T|I) < \sum_{i=1}^3 C_i'(T|I)$. \hfill $\Box$

\noindent \textbf{Note:} In the above proof, interference models other than the Protocol Model can be used as well if they are equivalent to the Protocol Model for the particular network considered there. The Interference Model~\cite{Kumar00} is one of them.

We next show that if $m=c$, the transport capacity of any Arbitrary Network is separable in channels.
\begin{theorem}
For an Arbitrary Network, if the number of radio
interfaces at each node is equal to the number of channels, i.e., if $m=c$, the separability property holds, i.e., $C(T) = \sum_{i=1}^c C_i'(T)$ as $T \to \infty$.
\label{theorem_sep}\end{theorem}
\noindent \textbf{Proof:} (a) We first show that $C(T) \geq \sum
C_i'(T)$. At first glance, it seems that any combined
configuration $G_1' \times ... \times G_c'$ is a special case of
$G$. However, there are two subtle difficulties here. The first
one is the interface location constraint mentioned before. With any feasible $G$, the interfaces of the
same node must have the same location, which is not guaranteed if
the interface locations of single-channel single-radio networks
$\mathcal{N}_i'$ are optimized independently. The second is the
source-destination pair selection. If the selections are done
independently, the same node may select different destinations on
$\mathcal{N}_l'$ and $\mathcal{N}_k'$ for $l\neq k$. The difficulties are
resolved by noting that the only difference between networks
$\mathcal{N}_i'$ is in the data rates $w_i$. Thus, any sequence of
STS's that occurred in one network $\mathcal{N}_l'$ can occur in
the same order in any other network $\mathcal{N}_k'$, and the only
difference is in the pace (proportional to $w_i$) at which the sequences occur. Therefore,
the optimal configurations $G_l'$ and $G_k'$ are the same except
for a constant scaling factor in time, making $G_1'^* \times ...
\times G_c'^*$ a special case of $G$. By the definition of $C(T)$,
we have
\begin{equation}
C(T) \geq \sum_{i=1}^c R(G_i'^*, T) = \sum_{i=1}^c C_i'(T). \label{eq_3a}
\end{equation}

(b) We now show $C(T) \leq \sum C_i'(T)$ as $T \to \infty$ by
showing that any sequence of \emph{STS}'s of network $\mathcal{N}$ 
can be \emph{simulated} by networks $\mathcal{N}_i'$, $i=1,...c$,
essentially in the same amount of time. Partition $T$ into
disjoint intervals $T_j$
\begin{equation}
T_j = \frac{w_j}{\sum_{i=1}^c w_i}T, ~~j=1,...,c. \label{eq_Tj}
\end{equation}
For any configuration $G$, it is clear that $\sum_{b \in \langle T
\rangle}l_b(G) = \sum_{j=1}^c \sum_{b \in \langle T_j
\rangle}l_b(G)$.

\begin{figure}[t]
\centerline{\psfig{figure=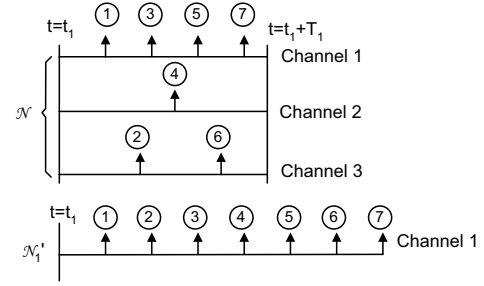,width=2.5in} }
\caption{Single-channel single-radio network $\mathcal{N}_1'$
simulates all 7 STS's completed on multi-channel multi-radio
network $\mathcal{N}$ in time interval $T_1$. A circled number
represents an STS, and an arrow indicates the time when an STS is
completed.} \label{fig_simulate}
\end{figure}

The simulation scheme is as follows: network $\mathcal{N}_j'$
simulates the STS's that occurred on network $\mathcal{N}$ during
time interval $T_j$, $j=1,...,c$, and the $c$ simulations run
simultaneously. That is, networks $\mathcal{N}_j'$ each replicate
a segment of the history of $\mathcal{N}$ in parallel. 

There is an important dependence among the STS's of network $\mathcal{N}$. Consider a bit that is forwarded by one hop in the current STS. This hop of forwarding contributes to the transport capacity only if the previous STS's have completed the previous hops of forwarding. To preserve this dependence, network $\mathcal{N}_j'$ schedules the STS's in the same order they were completed
in network $\mathcal{N}$ in $T_j$. This is illustrated in
Fig.~\ref{fig_simulate}. In network $\mathcal{N}$ during $T_1$, 4
STS's (numbered 1, 3, 5, 7) occurred on channel 1, 1 STS (numbered
4) on channel 2, and 2 STS's (numbered 2, 6) on channel 3. The
numeric order sequences the time instants at which the STS's were
completed on network $\mathcal{N}$. $\mathcal{N}_1'$ simulates all
7 STS's according to the numeric order. This way, $\mathcal{N}_1'$
delivers whatever bits that were delivered by network
$\mathcal{N}$ during $T_1$ and preserves the dependence among those bits. In general, the schedule is obtained
as follows. Define $L_{ij}:= \lceil T_j/\tau_i \rceil$, the number
of STS's completed on channel $i$ during $T_j$. Define $\Lambda_j := \{t
| t = \sum_{l=1}^{j-1}T_l + k_i \tau_i, i=1,...,c, k_i=1, ...,
L_{ij}\}$, $j=1,...,c$. That is, $\Lambda_j$ includes all the time
instants at which the STS's were completed on network
$\mathcal{N}$ during $T_j$. Sorting $\Lambda_j$ in ascending order forms
the schedule for network $\mathcal{N}_j'$.

To simulate the deliveries in $\langle T_j \rangle$, it takes
network $\mathcal{N}_j'$ time
\begin{eqnarray}
s_j & = & \sum_{i=1}^c \frac{w_i}{w_j}L_{ij}\tau_i
\\
& < & \sum_{i=1}^c \frac{w_i}{w_j} \left( \frac{T_j}{\tau_i} + 1 \right)\tau_i \\
& = & T + c \tau_j \label{eq_a1}
\end{eqnarray}
where (\ref{eq_a1}) follows from (\ref{eq_Tj}) and that $\tau_i =
1/w_i$. Also, note that $s_j \geq T$ because $L_{ij} = \lceil T_j/\tau_i \rceil \geq T_j/\tau_i$. We
obtain the following bounds
\begin{equation}
T \leq s_j < T + c \tau_j \label{eq_total_time}.
\end{equation}
Define $\hat{s} := \max_j \{ s_j \}$. Then $T \leq \hat{s} < T + c \max_j \{ \tau_j \}$. After an elapse of $\hat{s}$,
bit-meters in the amount of $\sum_{j=1}^c \sum_{b \in \langle T_j \rangle}l_b(G)
= \sum_{b \in \langle T \rangle}l_b(G)$ are achieved collectively
by networks $\mathcal{N}_j'$, and the transport rate is
\begin{eqnarray}
\hat{R} &= & \frac{\sum_{b \in \langle T \rangle}l_b(G)}{\hat{s}}\\
   &=& \frac{\sum_{b \in \langle T \rangle}l_b(G)}{T} \frac{T}{\hat{s}} \to
   R(G, T)
   \mbox { as } T \to \infty \label{eq_R'1}
\end{eqnarray}
On the other hand,
\begin{eqnarray}
\hat{R} &= & \frac{\sum_{j=1}^c \sum_{b \in \langle T_j \rangle}l_b(G)}{\hat{s}}\\
   & \leq & \sum_{j=1}^c \frac{\sum_{b \in \langle T_j \rangle}l_b(G)}{s_j} \\
   & \leq & \sum_{j=1}^c C_j'(T), \mbox{ as }T \to \infty \label{eq_use_def_c}
\end{eqnarray}
where (\ref{eq_use_def_c}) follows from the definition of
$C_j'(T)$. Combining (\ref{eq_use_def_c}) and (\ref{eq_R'1}) gives
\begin{equation}
R(G, T) \leq \sum_{j=1}^c C_j'(T), \forall G, \mbox{ as }T \to
\infty
\end{equation}
and hence, by the definition of $C(T)$,
\begin{equation}
C(T) \leq \sum_{j=1}^c C_j'(T), \mbox{ as }T \to \infty. \label{eq_3b}
\end{equation}

By (a) and (b), we have $C(T) = \sum_j C_j'(T)$, as $T \to \infty$.
\hfill $\Box$

\noindent \textbf{Note:} If we use a different simulation scheme, we may run into the following difficulty. When the multi-channel multi-radio network routes a bit through different channels, the transmissions simulated on $\mathcal{N}_i'$  may be disconnected, and thus do not contribute to the transport capacity $C_i'(T)$, which must be solely evaluated on $\mathcal{N}_i'$.

So far, we have considered $c+1$ networks: a multi-channel
multi-radio network, and the $c$ corresponding single-channel single-radio networks.
We next consider only one network (a multi-channel multi-radio
network) but two routing schemes.
\begin{theorem}
For an Arbitrary Network, given a channel assignment $I$, let the transport capacity under multi-channel routing be $C_{\mbox{mr}}(T|I)$, and let the transport capacity under single-channel routing be $C_{\mbox{sr}}(T|I)$. Then (1) $C_{\mbox{mr}}(T|I) \geq C_{\mbox{sr}}(T|I)$, $\forall \mathcal{N}$ and $I$, (2) $\exists \mathcal{N}$ and $I$ such that $C_{\mbox{mr}}(T|I) >
C_{\mbox{sr}}(T|I)$, and (3) if $m=c$, the two routing schemes result in equal capacities, i.e., $C_{\mbox{mr}}(T) = C_{\mbox{sr}}(T)$ as $T \to \infty$. \label{theorem_rt}
\end{theorem}
\noindent \textbf{Proof:} (1) This is true because single-channel routing is a special case of multi-channel routing.

\begin{figure}[t]
\centerline{\psfig{figure=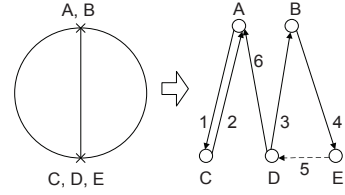,width=1.8in} }
\caption{The conditional optimal configuration, where nodes A, B are collocated at one end of a diameter, nodes C, D, E collocated at the other end of the diameter, and the numbers indicate the channels.}\label{mr_sr_arbi_fig}.
\end{figure}

(2) This is true because with multi-channel routing, some connected links that are on different channels can be used to deliver extra bits. Consider network $\mathcal{N}$ consisting of $n=5$ nodes $A$, $B$, $C$, $D$ and $E$, $\sigma$ being a circular disk, $m=3$, and $c=9$. The channel assignment $I$, in the form of ``(node: list of channels to which that node's interfaces are assigned)'', is ($A$: 1,2,6), ($B$:3,4,7), ($C$:1,2,8), ($D$:3,5,6), ($E$:4,5,9). The data rates of the channels are $w_i=2, i=1,...,4$, and $w_i=1, i=5,...,9$. Given $I$, the optimal configuration is shown in Fig.~\ref{mr_sr_arbi_fig} and is justified as follows. Node C must choose node A as its destination, since node A is the only node with which node C can communicate. Node A can communicate with node C on channel 1 or channel 2 and node D on channel 6. Since data rates $w_1,w_2 > w_6$, node A should choose node C as its destination. To maximize the contribution to the transport capacity, node A and node C must be at the opposite ends of a diameter, say $d_1$. Node D can communicate with node B on channel 3 and with node A on channel 6. Since $w_3 > w_6$, node D should choose node B as its destination. To maximize the contribution to the transport capacity, node D and node B must be at the opposite ends of another diameter $d_2$. Now since the only remaining idle link at node B is B$\leftrightarrow$E on channel 4, node B should choose node E as its destination. To maximize the contribution to the transport capacity, node E must be at the end of diameter $d_2$ that is opposite node B, and thus node E is collocated with node D.

Note that under multi-channel routing, node E, node D and node A form a path E$\to$D$\to$A traversing channel 5 and channel 6. Since node E and node D are collocated, node E should choose node A as its destination. To maximum the contribution to the transport capacity, node A must be at the end of diameter $d_2$ that is opposite node E. Note that node A is also on diameter $d_1$. Thus diameter $d_1$ and diameter $d_2$ overlap. The path E$\to$D$\to$A, which makes a positive contribution to the transport capacity, is forbidden under single-channel routing. Therefore, for network $\mathcal{N}$, multi-channel routing yields a higher transport capacity, i.e., $C_{\mbox{mr}}(T|I) >
C_{\mbox{sr}}(T|I)$.

(3) Now if $m=c$, we view all the interfaces on channel $j$ as a
single-channel single-radio network $\mathcal{N}_j$. Then, by
following part (b) of the proof of Theorem \ref{theorem_sep}, the
STS's under multi-channel routing can be simulated by the
$c$ networks $\mathcal{N}_j$ in parallel. Thus, $C_{\mbox{mr}}(T)
\leq \sum_{j=1}^c C_j(T) = C_{\mbox{sr}}(T)$, as $T \to \infty$,
which together with $C_{\mbox{mr}}(T) \geq C_{\mbox{sr}}(T)$
completes the proof. \hfill $\Box$

\section{Results for Random Networks}\label{sec_rand}
In a Random Network~\cite{Kumar00}, nodes are randomly located in a region. Each node randomly chooses another node as its destination, and as a result there are $n$ traffic flows in the network. The measure
of network capacity is the throughput capacity~\cite{Kumar00}, which is in the minimum sense since according to \cite{Kumar00} a throughput is defined to be feasible (i.e., to be admitted in evaluating the order behavior) if all nodes can achieve it. In some scenarios, it might be beneficial to consider the average of the actual throughputs of all nodes, and this prompts us to define the average-sense (AS) throughput capacity. We call the one in \cite{Kumar00} the minimum-sense (MS) throughput capacity. For the purpose of clarity, we give the definitions of both. But first we define the \emph{throughput}
of a flow originating from node $j$ during time interval $T$
\begin{equation}
\lambda(G, T, j) := N(G,T,j)/T,
\end{equation}
with unit bits/second, where configuration $G=(X, I, F, M, P)$ is defined in Section \ref{sec_not} and $N(G, T, j)$ is the number of bits delivered by the flow originating from node $j$ in a duration of $T$. Now we define the
\emph{minimum-sense (MS) throughput rate} under $G$ as
\begin{equation}
R(G, T) := n \min_{j=1,...,n} \lambda(G, T,j),
\end{equation}
where $n$ factors in the network size, and define the \emph{minimum-sense (MS) throughput capacity} as
\begin{equation}
C(T) := \mathrm{E}_{X,F} \sup_{G |(X, F)} R(G, T), \label{eq_cap_rand_no_cond}
\end{equation}
where $\mathrm{E}_{X,F}$ means taking the expected value with respect to $X$ and $F$, which are both uniform in their respective domains. We can define various conditional throughput capacities. For example, we define \emph{minimum sense (MS) throughput capacity conditioned on channel assignment $I$} as
\begin{equation}
C(T|I) := \mathrm{E}_{X,F} \sup_{G |(X, F, I)} R(G, T). \label{eq_cap_rand}
\end{equation}

We define the \emph{average-sense (AS) throughput rate} as
\begin{equation}
R(G, T) := \sum_{j=1}^n \lambda(G, T, j). \label{eq_rate_rand_avg}
\end{equation}
As in the minimum sense case, we can also define the \emph{average-sense (AS) throughput capacity} and the the \emph{average-sense (AS) throughput capacity conditioned on a channel assignment}, but we leave them out for brevity. It is clear that the AS throughput capacity is not less than the MS throughput capacity. As in Section \ref{sec_arb}, the corresponding single-channel
single-radio networks $\mathcal{N}_j'$ and their various throughput capacities can be defined as well.

\begin{theorem}
For a Random Network, given a channel assignment, the throughput capacity in general is not separable in channels, regardless of whether the throughput capacity is in the minimum sense or in the average sense.\label{theorem_sep_rd}
\end{theorem}
\noindent \textbf{Proof:} This is true because a flow in a single-channel single-radio network tends to have fewer hops and consequently a transmission may contribute more to the throughput capacity than a transmission in a multi-channel multi-radio network does. We prove this by showing the existence of a network $\mathcal{N}$ whose throughput capacity conditioned on a channel assignment $I$ is not separable in channels. The network $\mathcal{N}$ consists of $n=4$ nodes A, B, C and D, $m=2$, $c=4$, $w_i=1$ bits/sec where $i=1,...,4$, and the channel assignment $I$ is $(A:1,2)$, $(B:2,3)$, $(C:3,4)$, $(D:4,1)$.

We first consider the MS throughput capacity. Under channel assignment $I$, each of the corresponding single-channel single-radio networks $\mathcal{N}_i$ consists of two radio interfaces, and has a MS throughput capacity of $1$ bits/sec since $\sup_{0 \leq x \leq 1}  2\min \{x, 1-x \}=1$. Therefore, $\sum_{i=1}^c C_i'(T|I)=4$ bits/sec.

In network $\mathcal{N}$, each node can choose one of the three other nodes as its destination, resulting in $3^4=81$ flow configurations. Let the throughput capacity conditioned on $I$ and flow configuration $F$ be $C(T|I, F)$. It  can be checked that $C(T|I, F)$ is maximized if $F$ is either $F^\alpha$ = (A$\to$B, B$\to$C, C$\to$D, D$\to$A) or the reverse $F^\beta=$(B$\to$A, C$\to$B, D$\to$C, A$\to$D), each of which results in a $C(T|I, F)$ of $4$ bits/sec. Now consider another flow configuration: $F^\gamma=$(A$\to$B$\to$C, B$\to$C, C$\to$D, D$\to$A), which occurs with probability $1/81$. Let the throughput of flow A$\to$B$\to$C be $x$ bits/sec, where $0 \leq x \leq 1$, then the throughput of flow B$\to$C is $1-x$. The minimum of $x$ and $1-x$ is $0.5$ attained at $x=0.5$. The throughputs of the other two flows C$\to$D, D$\to$A are both $1$ bits/sec. Thus the minimum of the throughputs is $0.5$ bits/sec, and $C(T|I, F^\gamma) = 4\times 0.5=2$ bits/ec. Therefore,
$C(T|I) = \mathrm{E}_F C(T|I,F) \leq (1/81)2 + (1-1/81)4 < 4 = \sum_{i=1}^c C_i'(T|I)$.

We now consider the AS throughput capacity. It can be checked that $\sum_{i=1}^c C_i'(T|I)=4$ bits/sec. Also, $C(T|I,F)$ is maximized at flow configurations $F^\alpha$ and $F^\beta$ defined above. Now consider flow configurations: $F^\delta$ = (A$\to$B$\to$C, B$\to$C$\to$D, C$\to$D$\to$A, D$\to$A$\to$B), each flow having two hops. Due to symmetry, each flow has a throughput of $0.5$ bits/sec, and $C(T|I,F^\delta) = 4\times 0.5=2$ bits/ec. Following the argument in the case of MS throughput capacity, we have $C(T|I) < \sum_{i=1}^c C_i'(T|I)$. \hfill $\Box$

We next show that if $m=c$, the throughput capacity of any Random Network is separable in channels.
\begin{theorem}
For a Random Network, if $m=c$, the throughput capacity is separable in channels, i.e., $C(T) = \sum_{i=1}^c C_i'(T)$ as $T \to \infty$.
\end{theorem}
\noindent \textbf{Proof:} Fix $X$ and $F$, and set the distance of each hop to $1$, by the proof of Theorem \ref{theorem_sep}, we have $C(T|X,F)=\sum_{i=1}^c C_i'(T|X,F)$. Taking the expectation over $X$ and $F$ completes the proof. \hfill $\Box$

\begin{theorem}
For a Random Network, given a channel assignment $I$, let the throughput capacity under multi-channel routing be $C_{\mbox{mr}}(T|I)$, and let the throughput capacity under single-channel routing be $C_{\mbox{sr}}(T|I)$. Then (1) $C_{\mbox{mr}}(T|I) \geq C_{\mbox{sr}}(T|I)$, $\forall \mathcal{N}$ and $I$, (2) $\exists \mathcal{N}$ and $I$ such that $C_{\mbox{mr}}(T|I) > C_{\mbox{sr}}(T|I)$, and (3) if $m=c$, the two routing schemes result in equal capacities, i.e., $C_{\mbox{mr}}(T) = C_{\mbox{sr}}(T)$ as $T \to \infty$. \label{theorem_rt_rd}
\end{theorem}

\noindent \textbf{Proof:} (1) This is true because single-channel routing is a special case of multi-channel routing.

(2) We first prove the result for the MS throughput capacity. The result is true because the definition of MS throughput capacity may penalize single channel routing. Consider network $\mathcal{N}$ consisting of 3 nodes $A$, $B$ and $C$, $m=2$, $c=4$. The channel assignment $I$ is ($A$: 1,2), ($B$:2,3), and ($C$:3,4). There are 8 possible flow configurations, since each node can choose one of the two other nodes as its destination. Under single-channel routing, among those flow configurations, ($A \to B$, $B \to A$, $C \to B$) and ($A \to B$, $B \to C$, $C \to B$) have conditional throughput capacity $C(T|I,F) = 1/2$, and the other flow configurations have a conditional throughput capacity of $0$. Thus, $C_{\mbox{sr}}(T|I)=\frac{1}{8}\frac{1}{2}+ \frac{1}{8}\frac{1}{2} + 0= 1/8$. Under multi-channel routing, the first two flow configurations have the same conditional throughput capacity as they do under single-channel routing. Consider a third flow configuration ($A \to B$, $B \to C$, $C \to A$), which has a throughput capacity of $1/2>0$. Thus, $C_{\mbox{mr}}(T|I) > C_{\mbox{sr}}(T|I)$.

We now prove it for the AS throughput capacity. The result is true because with multi-channel routing, some connected links that are on different channels can be used to deliver extra bits. Refer to Table \ref{table_rand_mr_as} for the network $\mathcal{N}$ under consideration. Network $\mathcal{N}$consists of 5 nodes $A$, $B$, $C$, $D$ and $E$, $m=2$, $c=4$, $w_1=1$ bits/s, $w_2=6$ bits/s, $w_3=10$ bits/s, $w_4=1$ bits/s. The channel assignment $I$ is ($A$: 1,2), ($B$:2,3), ($C$:3,4), ($D$:1,3), and ($E$:1,4). For flow configuration $F^\alpha=$($A \to C$, $C \to A$, $D \to A$, $B \to E$, $E \to D$), under single-channel routing, the throughputs are $0$, $0$, $\leq 1$, $0$, and $\leq 1$, respectively, resulting in $C_{\mbox{sr}}(T|I, F^\alpha)\leq 2$. Under multi-channel routing, consider the following routing scheme: ($A  \stackrel{2}{\to} B \stackrel{3}{\to} C$, $C \stackrel{3}{\to} B \stackrel{2}{\to} A$, $D \stackrel{3}{\to} B \stackrel{2}{\to} A$, $B \stackrel{3}{\to} C \stackrel{4}{\to} E$, $E \stackrel{1}{\to} D$). The first three flows have an aggregate throughput of $6$, and the remaining two both have $1$, resulting in an aggregate throughput of $8$ and hence $C_{\mbox{mr}}(T|I, F^\alpha) \geq 8 > C_{\mbox{sr}}(T|I, F^\alpha)$. For any other flow configuration $F^\beta$, $C_{\mbox{mr}}(T|I, F^\beta) \geq C_{\mbox{sr}}(T|I, F^\beta)$. Taking the expected value of flow configuration completes the proof.

(3) If $m=c$, fix $X$ and $F$, and set the distance of each hop to $1$, by the proof of Theorem \ref{theorem_rt}, we have $C_{\mbox{mr}}(T|X,F) =C_{\mbox{sr}}(T|X,F)$, as $T \to \infty$. Taking the expectation of
$X$ and $F$ completes the proof. \hfill $\Box$

\noindent \textbf{Note:} It can be shown that for the first network in part (2) of the above proof, the AS throughput capacity under two routing schemes are equal, which together with the proof demonstrates the difference between AS throughput capacity and MS throughput capacity.

\begin{table}\caption{Channel Assignment for the Proof of Part (2) of Theorem \ref{theorem_rt_rd}} \label{table_rand_mr_as}
\begin{center}
\begin{tabular}{|c|c||ccccc|}
\hline
 Channel & Data Rate & A       & B        & C        & D          & E        \\
 \hline\hline
 1       & 1         & $\times$  &          &          & $\times$ & $\times$ \\
 2       & 6         & $\times$  & $\times$ &          &          &          \\
 3       & 10        &           & $\times$ & $\times$ & $\times$ &          \\
 4       & 1         &           &          & $\times$ &          & $\times$ \\
 \hline
\end{tabular}
\end{center}
\end{table}

\section{Implications of the Results}\label{sec_implications}
The results of this paper apply to networks of any size, including practical networks, which have a limited number of nodes.

An implication of the separability property is that if we know the formula for calculating the capacity of single-channel single-radio networks, in general we cannot calculate the capacity of a multi-channel multi-radio network by adding up the capacities of the corresponding single-channel single-radio networks. However, if the number of interfaces at each node is equal to the number of channels, this calculation is correct.

Another implication of the separability property is that if the number of interfaces at each node is equal to the number of channels, the network capacity linearly increases with the number of interfaces at each node.

An implication of the results on multi-channel routing is that allowing a packet to be routed to a channel different from the channel on which the packet is received may improve the network capacity. However, if the number of interfaces at each node is equal to the number of channels, this routing scheme does not improve the network capacity.

This paper does not answer the question of how to achieve the network capacity. To provide the answer, a multi-commodity flow problem can be formulated~\cite{Kod05}. However, the solution is computationally hard and only heuristic solutions have been found.

\bibliographystyle{ieeetr}

\bibliography{../../Lib/lma}

\end{document}